\documentclass[aps,prl,showpacs,twocolumn,amsmath,amssymb,nofootinbib,superscriptaddress,floatfix]{revtex4-1}
\def\beq{\begin{equation}}
\def\eeq{\end{equation}}
\def\bra#1{\left< #1\right|}
\def\ket#1{\left| #1\right>}
\def\bracket#1#2{\left<#1\mid #2\right>}
\def\EV#1#2#3{\bra{#1}#2\ket{#3}}

\usepackage[hypertex]{hyperref}
\usepackage{amsmath,amssymb, wasysym} 
\usepackage{verbatim}
\usepackage{graphicx}
\usepackage{epsfig}
\usepackage[usenames]{color}
\usepackage{color}
\usepackage{srcltx}
\usepackage{subfigure}
\begin{document}

\title{Logarithmic terms in entanglement entropies of 2D quantum critical points and Shannon entropies of spin chains}

\author{Michael P. Zaletel}
\affiliation{Department of Physics, University of California, Berkeley, California 94720, USA}
\author{Jens H. Bardarson}
\affiliation{Department of Physics, University of California, Berkeley, California 94720, USA}
\affiliation{Materials Sciences Division, Lawrence Berkeley National Laboratory, Berkeley, CA 94720, USA}
\author{Joel E. Moore}
\affiliation{Department of Physics, University of California, Berkeley, California 94720, USA}
\affiliation{Materials Sciences Division, Lawrence Berkeley National Laboratory, Berkeley, CA 94720, USA}

\begin{abstract}
Universal logarithmic terms in the entanglement entropy appear at quantum critical points (QCPs) in one dimension (1D) and have been predicted in 2D at QCPs described by 2D conformal field theories.  The entanglement entropy in a strip geometry at such QCPs can be obtained via the  ``Shannon entropy'' of a 1D spin chain with open boundary conditions.  The Shannon entropy of the XXZ chain is found to have a logarithmic term that implies, for the QCP of the square-lattice quantum dimer model, a logarithm with universal coefficient $\pm 0.25$.  However, the logarithm in the Shannon entropy of the transverse-field Ising model, which corresponds to entanglement in the 2D Ising conformal QCP, is found to have a singular dependence on replica or R\'enyi index resulting from flows to different boundary conditions at the entanglement cut.
\end{abstract}


\maketitle

The use of quantum information concepts to understand many-particle systems near a quantum critical point (QCP) has grown rapidly since the 1994 calculation of entanglement entropy in one-dimensional (1D) critical systems~\cite{holzhey}.  Entanglement entropy shows a universal logarithmic divergence at 1D critical points described by 2D conformal field theories (CFTs)~\cite{Vidal03, *Calabrese04} and at infinite-randomness 1D critical points~\cite{refaelmoore,*laflorencie,*santachiara}.  Recent work has studied generalizations to R\'enyi entropy and the entanglement spectrum~\cite{Calabrese:2008p1439,*fagotti}.  Terms proportional to $\log L$ for a subsystem of size $L$ in an infinite background are especially important as their coefficients are independent of the microscopic lattice spacing and hence potentially universal.  Entanglement can be used to develop a classification of 1D interacting quantum systems~\cite{guwen} and determines the difficulty of numerical simulation of 1D QCPs via matrix product state methods~\cite{pollmann}.

Above one spatial dimension, there are few general results on critical entanglement entropy.  For critical points described by a $d+1$-dimensional CFT  there is a conjecture for some geometries from the AdS/CFT correspondence~\cite{Ryu06} for the powers of length that appear in the entanglement entropy: logarithmic terms appear generically only in odd spatial dimensions.  This conjecture agrees with 1D results, as well as perturbation theory~\cite{metlitskisachdev} and variational methods~\cite{tagliocozzo} above 1D.  For another class of systems, ``conformal quantum critical points'' (CQCPs) in $d=2$~\cite{ardonneff, *IsakovFendley}, whose ground state wavefunctions are related to 2D rather than 3D CFT's, nonperturbative analytical and numerical results are possible and are the focus of the present work. An example of a CQCP is the critical state of the square lattice quantum dimer model~\cite{rokhsarkivelson} originally introduced as a model for high-temperature superconductors.  

We investigate logarithmic terms in the von Neumann entropy $S$ and R\'enyi entropies $S_n$ for CQCPs associated with the compact boson (which has CFT central charge $c=1$) and the Ising model ($c = 1/2$).  One result is the first explicit observation of a universal entanglement entropy logarithm above 1D.  For the $c=1$ boson, the results confirm a conjecture from the initial study of entanglement at CQCPs~\cite{FradkinMoore}, which derived a formula for a universal, geometry-dependent logarithmic correction to the leading area law.  Recent work~\cite{Stephan, HsuFradkin2,Oshikawa} on order-unity ($O(1)$) terms called into question the validity of that formula as for the $O(1)$ terms there are subtle issues related to compactification that took several years and several groups to resolve.  We give an improved derivation in supplementary material explaining why subtle factors affecting the $O(1)$ terms are irrelevant for the logarithm.

However, for the Ising CFT we find that the coefficient of the logarithm is discontinuous as the R\'enyi index $n$ passes through $n =1$.  We explain this behavior analytically by arguing that the correct ``defect line'' used in the calculation of the entropy changes discontinuously with $n$ and is not the combination of one free field and $n-1$ Dirichlet fields as conjectured in Ref.~\onlinecite{FradkinMoore} and confirmed for the boson.  A similar but less singular discontinuity in $n$ was previously found numerically to exist for the $O(1)$ term~\cite{StephanIsing}.   Our numerical approach to logarithmic terms uses large-scale Time-Evolving Block Decimation (TEBD)~\cite{GVidal} calculations to implement the same mapping used there between entanglement entropy at 2D CQCPs and the Shannon entropy of 1D spin chains.  The compact boson case relevant to the quantum dimer model seems to be well understood, but the results here demonstrate that the correct defect boundary condition can be complicated and $n$-dependent.

\begin{figure}
\includegraphics[width=0.45\textwidth]{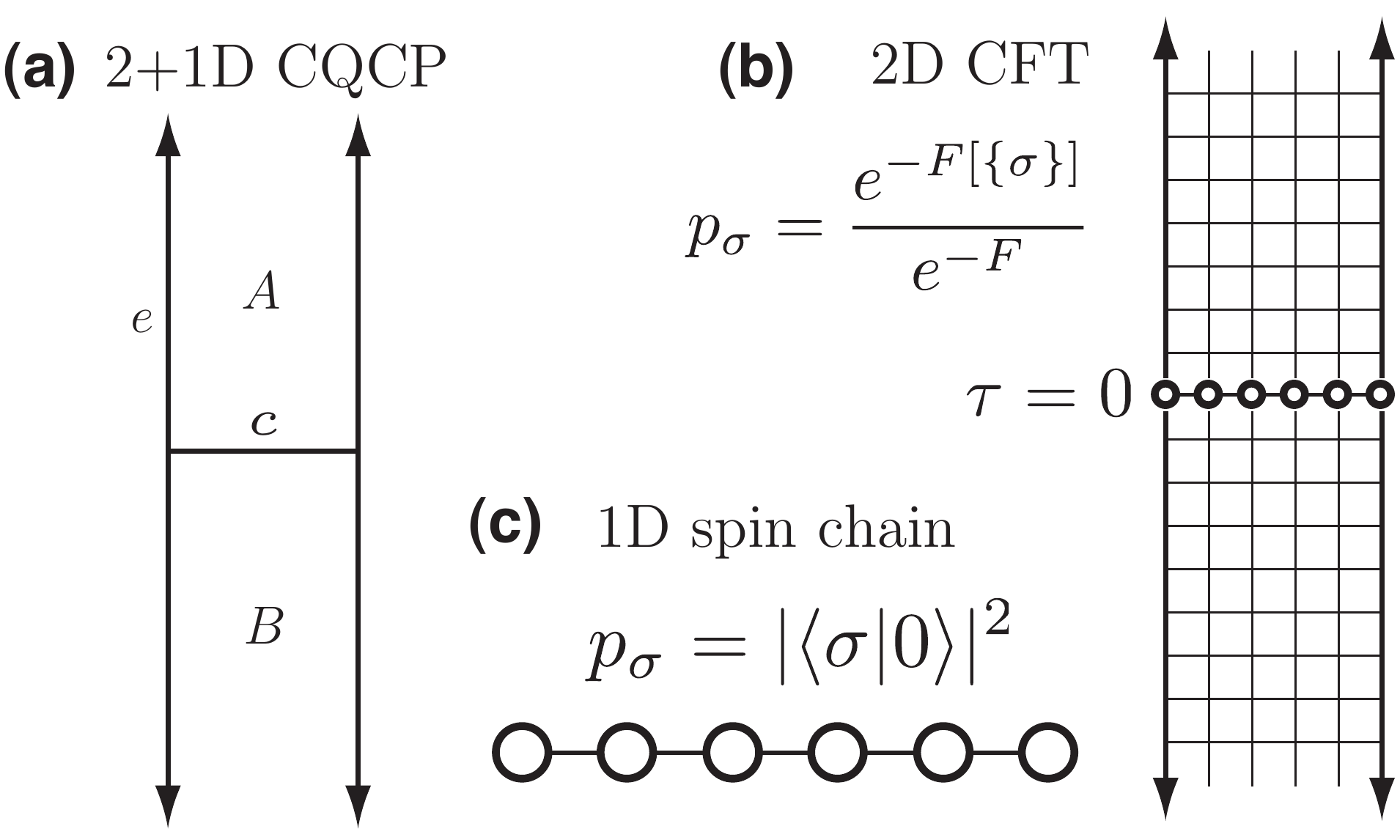}
\caption{ a) The entanglement spectrum of the CQCP, $p_\sigma$, in a strip geometry is in one-to-one correspondence with field configurations $\{ \sigma \} $ on the entanglement cut. These probabilities can be alternately understood as b), partition functions for a constrained CFT in the strip geometry, or c), ground state amplitudes of a quantum chain.  } \label{Transfer}
\end{figure}
	
We first define some basic concepts. The bipartite entanglement spectrum $\{ p_\sigma \}$ of a pure quantum state consists of the eigenvalues of the reduced density matrix of either subsystem.  The von Neumann entropy $S$ and R\'enyi entropies $S_n$ are defined by
\begin{equation} \label{RenyiDef}
S = - \sum_{ \left \{ \sigma \right \} } p_\sigma \log p_\sigma,  \quad S_n = \frac{1}{1 - n} \log( \sum_{ \left \{ \sigma \right \} } p_\sigma ^n ). 
\end{equation}

	The entanglement spectrum $\{ p_\sigma \}$ of a CQCP ground state in strip or cylinder geometries can be
  understood both in terms of the associated 2D CFT or as probability amplitudes of a 1+1D spin chain, as
  illustrated in Fig.~\ref{Transfer}. The entanglement spectrum $\{ p_\sigma \}$ is in one-to-one correspondence
  with CFT field configurations $\{ \sigma \}$ along the entanglement cut.  If the CQCP lies on a strip or
  cylinder, the 2D CFT naturally defines a 1D quantum Hamiltonian $H$ via the transfer matrix formalism. `Time' is
  chosen parallel to the strip (cylinder), with the entanglement cut at a fixed time $\tau = 0$. For universal
  properties, we can take the spatial direction to lie on a lattice. In the limit of an infinite strip the ground state probabilities are identical to the
  entanglement spectrum of the associated CQCP~\cite{Stephan}.  The Shannon entropy of the spin chain, defined
  as in Eq.~\eqref{RenyiDef} with $p_\sigma = |\langle \sigma|0\rangle|^2$, is a basis dependent measure of
  disorder in the ground state. The Shannon entropy of the spin chain is then equivalent
  to the bipartite von Neumann entropy of the CQCP.

	The relationship between the CQCP entanglement spectrum, spin chain ground state probabilities, and CFT partition function is summarized by	
\begin{align}
p_\sigma &= \lim_{\beta \to \infty} \frac{ \EV{i}{e^{- \beta H}}{\sigma} \EV{\sigma}{e^{- \beta H}}{i} }{\EV{i}{e^{-2 \beta H}}{i}} = |\bracket{\sigma}{0}|^2 \label{pQdef} \\
&= \lim_{m \to \infty} \frac{ \EV{i}{\mathcal{T}^m}{\sigma} \EV{\sigma}{\mathcal{T}^m}{i} }{\EV{i}{\mathcal{T}^{2 m}}{i}} = \frac{e^{-  F[ \{ \sigma \}]}}{e^{- F}} \, \, . \label{pSdef}
\end{align}
Here $\mathcal{T}$ is the transfer matrix of the CFT, and the result is independent of the boundary state $\ket{i}$ in the limit of an infinite strip. With a basis $\{ \sigma \}$ for the Shannon entropy, the projection operator $\ket{\sigma} \bra{\sigma}$ in the numerator of Eq.~\eqref{pSdef} constrains the corresponding field of the CFT along a cut at $\tau = 0$; the bulk fields remain free. The numerator and denominator are thus constrained and unconstrained partition functions, with free energies $F[\{ \sigma \}]$ and $F$ respectively. 

The R\'enyi entropies $S_n$ suggest a thermodynamic notation
\begin{align}
\mathcal{Z}(n) = e^{ - F(n) } &\equiv  \sum_{ \left \{ \sigma \right \} } p_\sigma^n = \sum_{ \left \{ \sigma \right \} } e^{- n \left( F[ \{ \sigma \}] - F \right) },
\end{align}
defined so that $F(n) = (n - 1) S_n$. The `free energy' $F(n)$ is found to depend on the length of the chain $L$ as
\begin{equation}
F(n, L) = f_1(n) L + \gamma(n) \log(L/a) + f_0(n) + \cdots \, \, \, .
\end{equation}
At a critical point $\gamma(n)$ should be universal and determined by the CFT as it is independent of the UV cutoff $1/a$. In the case of a cylinder geometry, $\gamma(n)$ is zero for an arbitrary CFT as the trace anomaly from a smooth conformal defect line is zero.  For a strip geometry, a universal logarithmic term is expected as explained below.

	For integer $n$, $\mathcal{Z}(n)$ can be interpreted as the partition function of a replicated CFT with $n$ copies of the field constrained to agree along $\tau = 0$, normalized by the free partition function. $F(n)$ is then interpreted as the free energy of a defect line in the replicated CFT. If the logarithmic contributions can be calculated in the replicated CFT and analytically continued to $n = 1$, then the entire spectrum of R\'enyi entropies $S_n$  as well as $S$ is known, via $S = \partial_n F(n)|_{n=1}$. (Below, this `replica trick' is found to fail for the Ising model due to non-analyticity of $F(n)$ at $n=1$ in the thermodynamic limit.)
		
We start with numerical evidence for the existence of logarithmic terms in the case of the $c=1$ compact boson, the CFT associated with the CQCP of the square-lattice quantum dimer model.  For a free boson, Ref.~\onlinecite{FradkinMoore} predicted a contribution $-\frac{1}{4} \log(L) \in S$ in the strip geometry; we indeed find a logarithm $- \frac{1}{4} \log(L)$ for the external boundary conditions (on the edge of the strip) assumed there, and for an alternate boundary condition we obtain $+ \frac{1}{4} \log(L)$ as derived below.
By using the transfer matrix mapping, it is sufficient to calculate the Shannon entropy of a 1D quantum model, the spin-1/2 XXZ chain
\begin{equation}
H = - h (\sigma^x_1 + \sigma^x_L ) + \sum_{i = 1}^{L-1} \left[ \sigma^x_i \sigma^x_{i+1} + \sigma^y_i \sigma^y_{i+1} + \Delta \sigma^z_i \sigma^z_{i+1} \right].
\end{equation}
In the continuum limit the XXZ model is described by a compact boson. The boundary conditions are tuned by an external field $h$ on the surface sites, and the Shannon entropy is calculated using the $\sigma^z$ basis. 

\begin{figure}[t]
\subfigure[]{\includegraphics[width=0.235\textwidth]{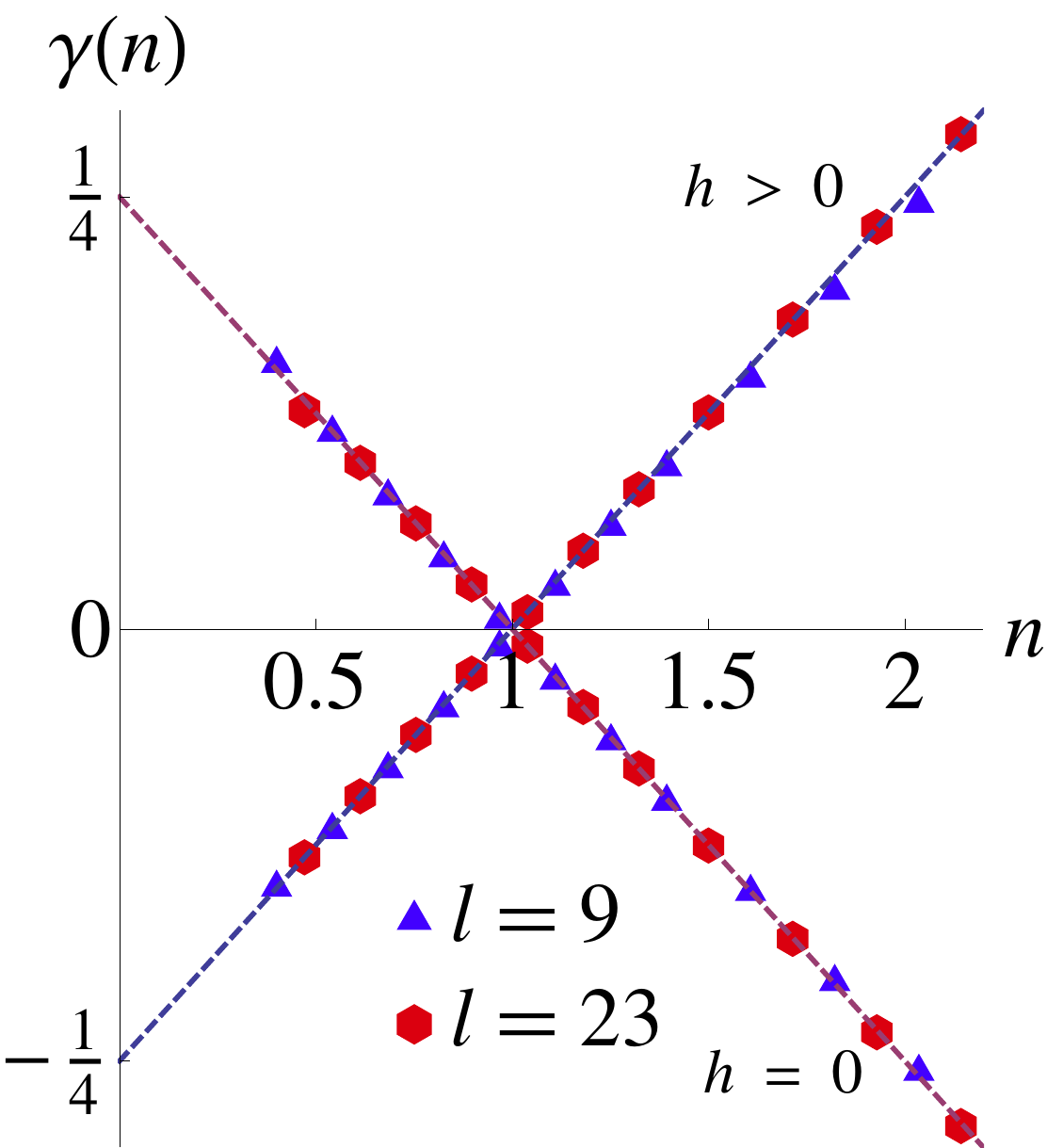}} 
\subfigure[]{\includegraphics[width=0.23\textwidth]{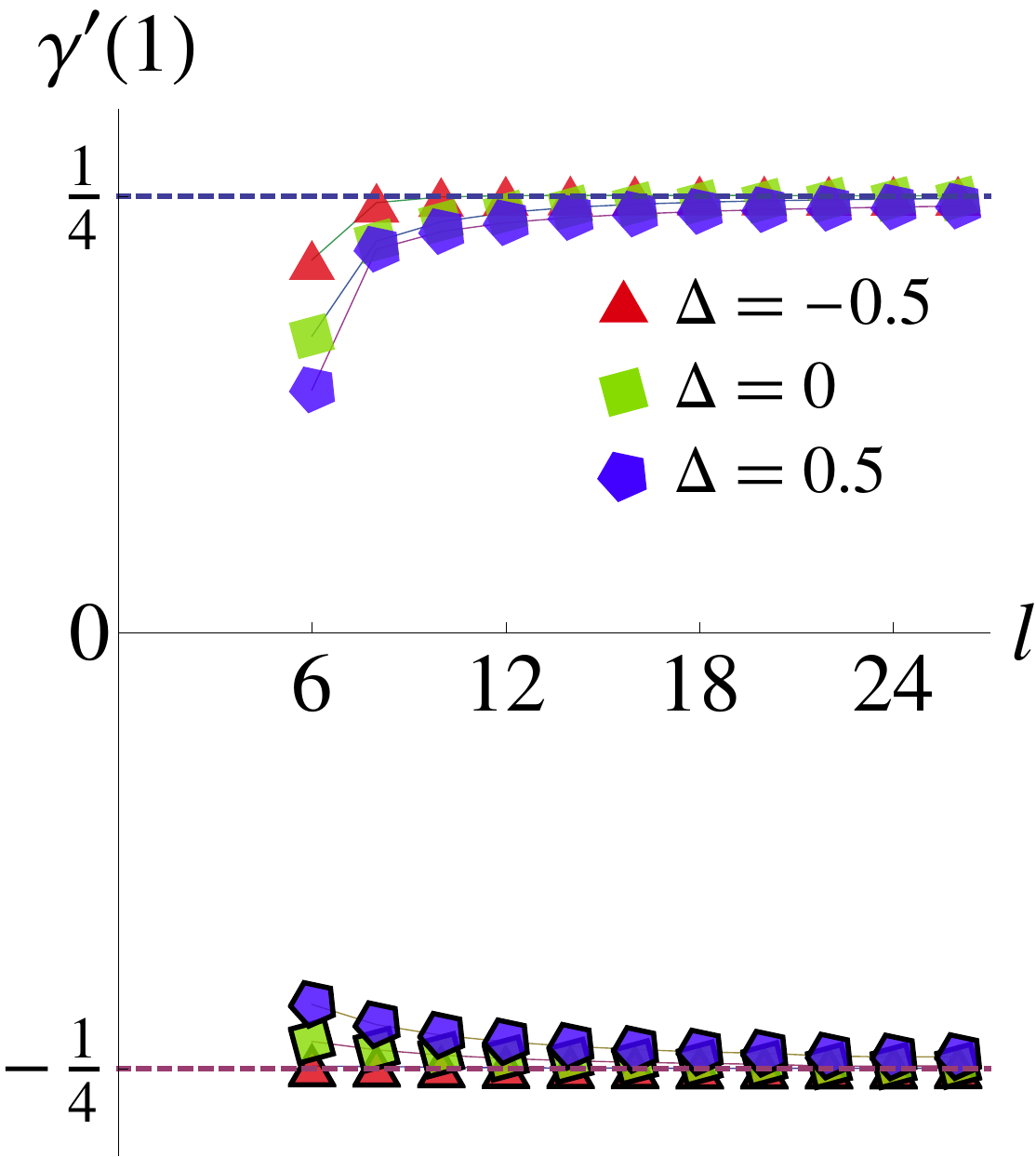}}
\caption{ a) The logarithmic part of $F(n)$ at $\Delta = - 1/2$. Data is included both without ($h = 0$) and with ($h > 0$) an edge magnetic field. $F(n, L)$ was computed for $L = 2, 4, \cdots 30$. Windows of 6 data points centered at $\ell$, $L = \{ \ell - 5, \cdots, \ell + 5 \}$, were fitted to the form $f_1 L + \gamma \log(L) + f_0 + f_{-1} L^{-1}.$ The data displayed includes $\gamma$ for $\ell = \{ 9, 23 \}$. The approach of $\gamma(n)$ to the dashed lines suggests convergence to $\gamma(n) = \mp \frac{1}{4} (n - 1)$ as $\ell \rightarrow \infty$, for $h= 0$ and $h > 0$ respectively. The inclusion of a $L^{-1}$ term accelerates convergence, but the value of $\gamma(n)$ remains insensitive to within a few percent to the choice of fitting form for terms that vanish as $L \rightarrow \infty$, as discussed in the supplementary material. b) The logarithmic contribution to the von Neumann entropy $S(L)$, as a function of the fitting window center $\ell$. As $\ell$ increases, the logarithm's coefficient converges to $\mp 1/4$, depending on the boundary conditions. Successive lines are for $\Delta = \{ -1/2, 0 , 1/2 \} $.}  \label{RenyiLogm2} \label{XXZVonN}
\end{figure}

	The ground state was found by TEBD for L = $\{2, 4, \cdots, 30 \}$ at $\Delta =  \{ -\frac{1}{2}, 0, \frac{1}{4}, \frac{1}{2}, \frac{3}{4} \}$. We estimate the ground state error  to be of order $\epsilon  = 1 - |\bracket{0}{\Psi}|^2 \sim \mathcal{O}(10^{-6})$, which was verified for the exactly solved $\Delta = 0$ case. The partition function $\mathcal{Z}(n)$ was then computed by summing over all $2^L$ configurations. The logarithmic contribution $\gamma(n)$ was extracted from the finite-size data as described in Fig.~\ref{RenyiLogm2}a. $\gamma(n)$ is well described by 	
\begin{equation} \label{XXZg}
\gamma(n) = \mp \frac{1}{4} (n - 1)
\end{equation}
for $h = 0$ and $h > 0$  boundary conditions respectively. For the compact free boson, Eq.~\eqref{XXZg} is correct for all $n$, while for XXZ and dimer lattice realizations, the result is modified in the regime $n \geq 2$ due to additional boundary operators becoming relevant~\cite{Stephan3}.

The slope of $\gamma(n)$ at $n=1$ is of special interest as the von Neumann entropy is given by $S = \partial_n F(n)|_{n=1}$ if the derivative exists.  In the case of the XXZ model, there is a logarithmic contribution $\partial_n \gamma(n)|_{n=1} =  \mp \frac{1}{4}$, so the von Neumann entropy scales as
\begin{equation}	\label{XXZS}
S(L) = s_1 \cdot L \mp \frac{1}{4} \log(L/a) + \cdots  \, \, \, .
\end{equation}
The rapid convergence of $S(L)$ to Eq.~\eqref{XXZS} with increasing $L$ is illustrated in Fig.~\ref{XXZVonN}b.

	To derive Eq.~\eqref{XXZg}, we perform a Jordan-Wigner transformation on the XXZ spin chain and bosonize the resulting fermionic model.  The model is mapped onto the universality class of a free, compact boson $\phi$ for $|\Delta| < 1$, with a compactification radius $R$ that depends on $\Delta$. For $h = 0$ the external boundary condition (b.c.) is Dirichlet in the continuum limit, $\phi = 0$, while for $h \neq 0$ the external b.c.\ is Neumann, $\partial_x \phi = 0$~\cite{Affleck}. The resulting change in the logarithmic contribution is thus attributed to `boundary condition changing operators' (bcc operators) in the underlying CFT. For integer $n$, the replicated theory contains $n$ copies of the field, $\phi^a$, subject to the constraint $\phi^a = \phi^b$ along the $\tau = 0 $ cut, modulo compactification.  In the analogous problem for a non-compact boson, the action, external b.c.'s, and path integral measure are all invariant under an $O(n)$ rotation of the replica index, so we can rotate the replicated fields to a new basis which includes the `center of mass' field $\phi_{CM} = \frac{1}{\sqrt{n}} \sum_{a=1}^n \phi^a$. The gluing condition now factorizes, giving the free center of mass field plus $(n-1)$ decoupled fields satisfying Dirichlet b.c. on the cut~\onlinecite{FradkinMoore}.  The free energy is
\begin{equation}
F(n) = (n-1) (F_D - F)
\end{equation}
where $F_D$ is the free energy of a \emph{single} replica with Dirichlet b.c. on the cut and $F$ is the unconstrained free energy. Using this rotation, the logarithmic contribution to $F(n)$ follows from the free energies $F_D$ and $F$ of a single replica.

	Logarithmic contributions to the free energy of a CFT arise from a `trace anomaly,' in which the trace of the stress tensor $\left< T^\mu_\mu \right>$ does not vanish. In the relevant `cut strip' geometry, the anomaly arises from the four corners where the cut along $\tau = 0$ terminates into the external boundary, as illustrated in Fig.~\ref{Transfer}a. The geometry of the corner contributes a term proportional to the central charge $c$~\cite{Cardy88}. However, there can also be a contribution due to the changing b.c.: if the external b.c. is Neumann, then termination of the Dirichlet b.c. on the cut into the external Neumann b.c.\ introduces a bcc operator~\cite{CardyBCFT}. The scaling dimension of this bcc operator, $h_{N D}$, will appear as an extra contribution to the trace anomaly.
	
For an arbitrary CFT in the `cut strip' geometry the predicted coefficient of the logarithmic term is
\begin{equation} \label{TraceAnomaly}
L \partial_L (F_c - F)  =  2 \times 2 \left[ 2 h_{e c} - \frac{c}{16} \right] \, .
\end{equation}
$F$ is the free energy of the strip with no constraint at $\tau = 0$, while $F_c$ is the free energy when constrained to some b.c.\ `$c$' along the cut, and both obey external b.c.\ `$e$'. Here $h_{e c}$ is the scaling dimension of the bcc operator required to take the external b.c.\ $e$ to the b.c.\ $c$ on the cut. This result can be extended if e.g. the cut is not perpendicular to the boundary. 
	In the compact boson case, $c = 1$, and the scaling dimension $h_{e D}$ is $0$ or $1/16$ for $e = D, N$ respectively. Substituting into Eq.~\eqref{TraceAnomaly}, $ L \partial_L (F_D - F)  =  \mp \frac{1}{4}$ for $D/N$ respectively, i.e., the sign of the logarithmic term depends on the external b.c.
The XXZ results are well described by the rotation trick and this bcc operator effect.
It can be shown (supplementary material) that the logarithm is unaffected by compactification, as supported by the XXZ data. Briefly, field configurations can be decomposed into two sectors: fluctuations versus zero modes or vortices. Only the latter, topological sector is affected by compactification. The fluctuating sector can be rotated.  The topological sector has exact conformal invariance: the trace anomaly and logarithm arise only from the fluctuations. Hence the logarithmic terms are compactification-independent while the $O(1)$ terms, arising from both sectors, are not.

The $c=1/2$ Ising model exhibits strikingly different behavior from the compact boson. In the transfer matrix formalism, it is sufficient to consider the Shannon entropy of the critical transverse-field Ising model,	
\begin{equation}
H = - \sum_{i = 1}^{N-1} \sigma^z_i \sigma^z_{i+1} + \sum_{i=1}^N \sigma^x_i  
\end{equation}
The Shannon entropy is calculated in the $\sigma^z$ basis.

\begin{figure}[t]
\includegraphics[width = 0.48\textwidth]{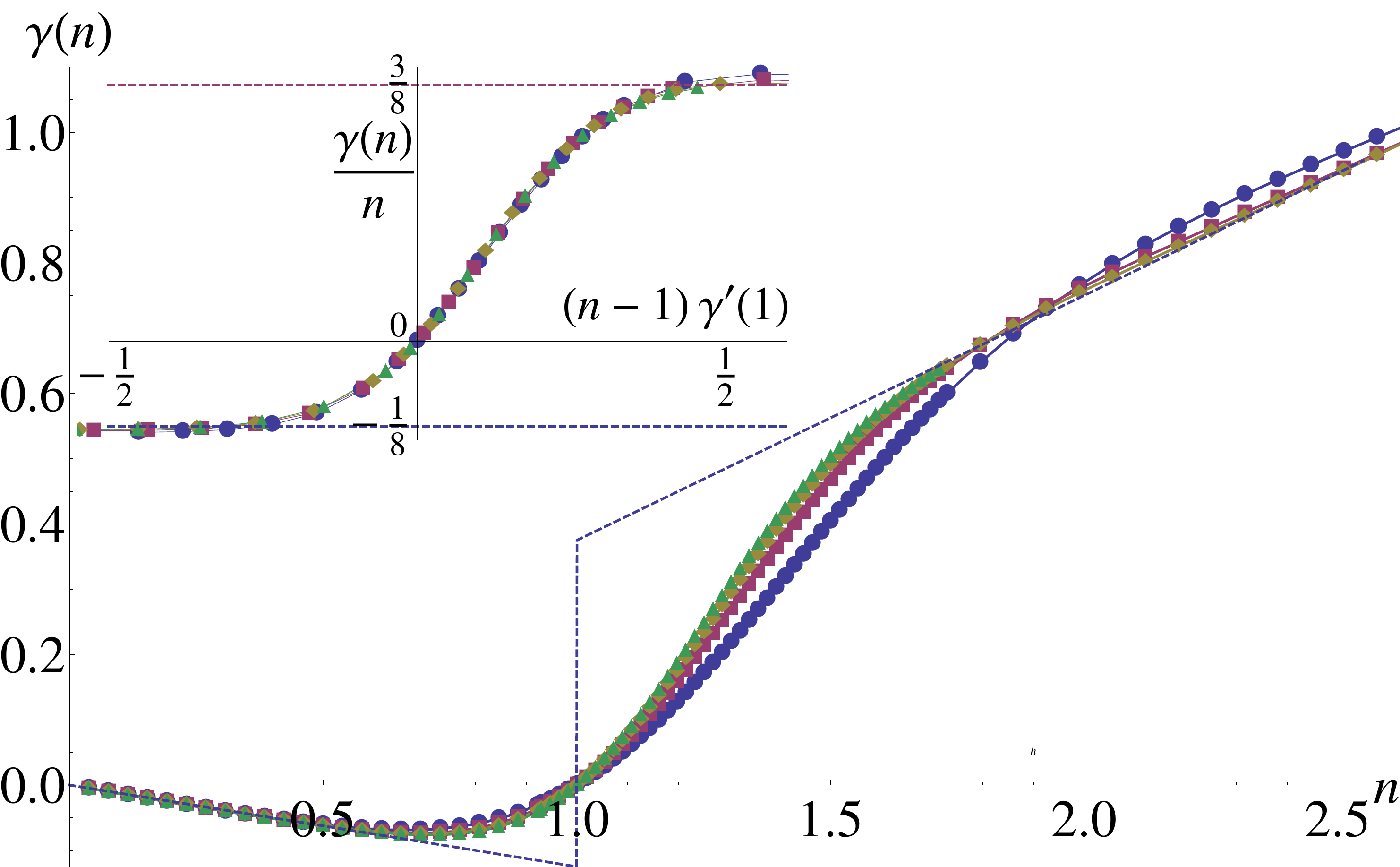}
\caption{ The logarithmic part of $F(n)$  for the TFI model. $F(n, L)$ was computed for $L = 2, 4, \cdots 40$, and $\gamma(n)$ extracted by the same procedure as for the XXZ model. The dashed lines are $\gamma(n) = -\frac{1}{8} n$ and $\gamma(n) = \frac{3}{8} n$. In the inset, the data is collapsed by plotting $\gamma(n)/n$ as a function of $(n-1)\gamma'(1)$.
}
\label{RenyiLogTFI}
\end{figure}
	The extracted logarithm $\gamma(n)$, shown in Fig.~\ref{RenyiLogTFI}, suggests a discontinuous $n$-dependence at $n = 1$. The dashed lines illustrate the likely convergence to
\begin{equation}
\gamma(n) =
\begin{cases}
-\frac{1}{8} n & \mbox{ for } n < 1 \\
0 & \mbox{ for } n = 1\\
\frac{3}{8} n & \mbox{ for } n > 1 \\
\end{cases} \, \, .
\end{equation}
Note that the rotation used in the boson case predicts that $\gamma(n) \propto (n - 1)$, as there is always one `free' center of mass field. In the present Ising case, $\gamma(n) \propto n$, which is evidence that the rotation trick is not applicable and a new analysis is required. In the inset of Fig.~\ref{RenyiLogTFI}, the data is collapsed by plotting $\gamma(n)/n$ as a function of $\gamma'(1)(n-1)$. Recall $\gamma(1) = 0$ identically due to the normalization of the entanglement spectrum.

	The large and small $n$ limits can be understood as follows. At large $n$, $\mathcal{Z}(n)$ is dominated by the probability $p_{\uparrow}  = |\bracket{\uparrow \uparrow \cdots}{0}|^2$ and its partner $p_\downarrow$. In this `low temperature' regime, the defect line at $\tau = 0$ is asymptotically a `fixed' defect line. Now all $n$ fields experience this fixed defect condition, not $n-1$ of them, and	
\begin{equation}
F(n) \sim n (F_{\uparrow} - F) - \log(2)
\end{equation}
Here $F_{\uparrow}$ is the free energy of a \emph{single} replica with spins constrained to $\sigma = +1$ along the defect. Again, there is an anomaly due to the `cut strip' geometry. The external boundary conditions are free, so the relevant scaling dimension is $h_{f \uparrow} = \frac{1}{16}$ \cite{CardyBCFT}. Using Eq.~\eqref{TraceAnomaly}, $ \gamma(n) = \frac{3}{8} n $ as observed. Likewise, in the small $n$ limit, the defect becomes disordered. This is obvious at the $n = 1/2$ point, where the geometry is in fact a half-strip with a free boundary at $\tau = 0$. With $h_{f f} = 0$, we arrive at $\gamma(n) =  - \frac{1}{8} n $ as observed.  While these results strictly apply only in large and small $n$ limits, the data appears to support an $n$-dependent phase transition: for $n < 1$, the defect flows to free boundary conditions for $2n$ decoupled half strips, while for $n > 1$, the defect is fixed. This interpretation is consistent with numerical results for the $O(1)$ term on a cylinder\cite{StephanIsing}. The renormalization group equations for these b.c. flows can be analyzed perturbatively~\cite{CardyRB}, and support the hypothesis of a phase transition. This perturbative approach could be generalized to other CFTs in order to study further possible entanglement boundary conditions and replica transitions.
	
The authors acknowledge conversations with I. Affleck, J. Cardy, E. Fradkin, A. Ludwig and M. Oshikawa, and support from DOE DE-AC02-05CH11231 (JHB) and NSF DMR-0804413 (JEM).

\clearpage
\onecolumngrid
\appendix
\setcounter{equation}{0}
\setcounter{page}{1}
\thispagestyle{empty}

\begin{center}
\large{ \textbf{Supplementary material for ``Logarithmic terms in entanglement entropies of 2D quantum critical points and Shannon entropies of spin chains''}}
\end{center}
\section{Fitting Procedure}
	In order to extract the logarithmic part of $F(n, L) = f_1(n) L + \gamma(n) \log(L) + f_0 + \cdots$ from finite size samples, $L = \{2, 4 , \cdots \}$, we performed a windowed fitting procedure: each window of 6 data points, $L = \{ \ell - 5, \ell - 3, \cdots, \ell + 5 \}$, was fit to the form $f_1(\ell, n) L + \gamma(n, \ell) \log(L) + f_0(n, \ell) + f_{-1}(n, \ell) L^{-1}$ using linear least squares. As $\ell$ increased, the coefficient $\gamma(n, \ell)$ of the XXZ model was observed to converge for the range of $n$ studied, $ 0.25 \lesssim n \lesssim 2$, as shown in Fig. 2 of the main text. Here we address the sensitivity of the result to the details of the fitting procedure.
		
\begin{figure}
\includegraphics[width=0.5\textwidth]{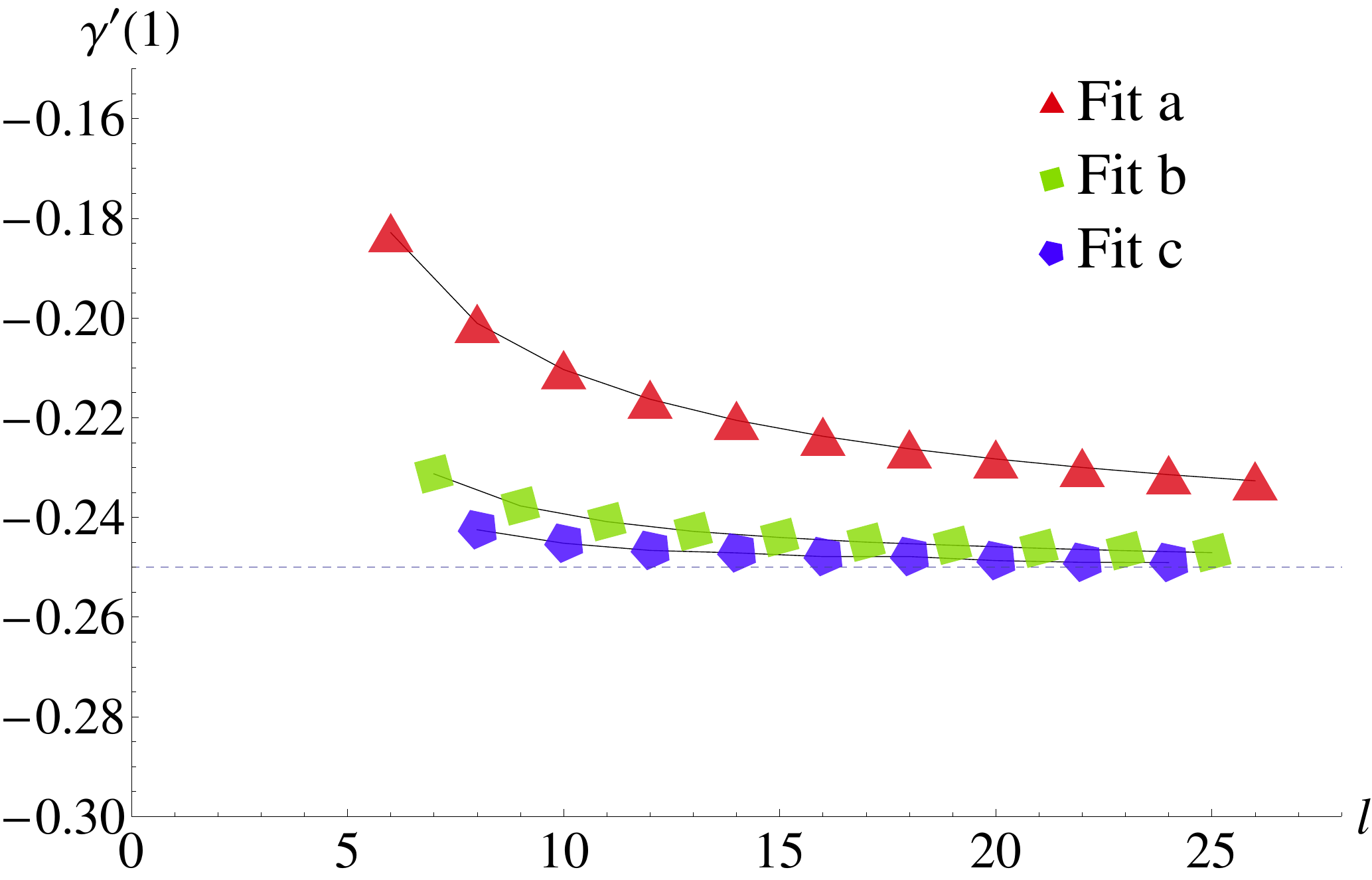}
\caption{ The extracted logarithm $\gamma'(n=1)$ as a function of fitting window, for three fitting procedures: a) No subleading terms, b) a subleading term $f_{-1} L^{-1}$, and c) subleading terms $f_{-1} L^{-1} + f_{-2} L^{-2}$. The data is taken from the XXZ model at $\Delta = \frac{1}{4}$ with Dirichlet boundary conditions.} \label{FittingDetail}
\end{figure}

		In principle, it is not necessary to include further subleading terms such as $L^{-1}$ if large enough system sizes are considered, but convergence can be accelerated by inclusion of terms arising from irrelevant operators. The choice of $L^{-1}$ is motivated by the scaling dimension of the stress tensor, which we generically expect to be present. To demonstrate the improved convergence, we perform the windowed fit with a) no terms beyond $f_0$, with a window size of 5, b) a subleading term $f_{-1} L^{-1}$, with a window size of 6 and c) subleading terms $f_{-1} L^{-1} + f_{-2} L^{-2}$, with a window size of 7. We extract $\gamma'( n = 1, \ell)$ using all three procedures at $\Delta = \frac{1}{4}$, with Dirichlet boundary conditions, as shown in Fig.~\ref{FittingDetail}. For the largest available window (with an upper bound of $L  = 30$), the three fitting procedures produce  $\gamma'(n = 1, \ell_{max}) = \{-0.233, -0.247, -0.249 \}$ for cases a), b), and c) respectively. In all three cases, the trajectory is consistent with convergence to $\gamma'(n = 1) = -0.25 \pm 0.01$. While including $L^{-1}$ and its descendants clearly accelerates the convergence, there are a multitude of other irrelevant operators, such as the vertex operators, with $n$ - dependent scaling dimensions. To test the sensitivity of the fit to the presence of different power law type terms, we replace $L^{-2}$ with a term $L^{-\alpha}$, with a 7 point window. We find that for $\alpha \in (0.5, 3)$, $\gamma'(n=1, \ell = 25)$ ranges between -0.248 and -0.256, with trajectories again consistent with $\gamma'(n=1) = 0.25 \pm 0.01$. To an accuracy of several percent, our conclusion is insensitive to the choice of subleading terms.

	A similar analysis holds for the Ising model, except near $n = 1$,  where the extracted logarithm is not yet close to converging for the system sizes considered. In the approximate region $0.8 < n < 1.25$,  the extracted logarithm is sensitive to the choice of subleading terms or the introduction of terms such as $\log(L)^2$. This is to be expected in any region far from convergence. However, a monotonic approach to the extrapolated values $\gamma(n) = \{ -\frac{1}{8} n, \frac{3}{8} n\}$ with increasing $\ell$ is a generic feature, as well as a slowly diverging derivative $\gamma'(1)$ at $n = 1$.

\section{R\'enyi Entropy of the Free Compact Boson CQCP}
The method applied by Ref.~\onlinecite{FradkinMooreS} to the non-compact boson does not strictly apply in the compact
case. Several works \cite{HsuFradkin1S, HsuFradkin2S, OshikawaS} have addressed the computation of the O(1) term in the case
of an infinite cylinder using the boundary conformal field theory of the compact boson. A recent preprint ~\cite{Stephan3S} computes the full partition function, again in the infinite cylinder or strip geometry, without using the replica trick, with the advantage of applying immediately to non-integer $n$. Here we evaluate the replicated partition function for an arbitrary geometry. We show that the logarithmic part is not changed by compactification, and give a formula for the O(1) contribution that can be extended to any geometry. Only for finite size subsystems is the effect of compactification really present, in the form of relative winding numbers which lead to Riemann Theta functions. In the limit of an infinite cylinder, we recover the formula for the O(1) term found in previous work.

	 We must evaluate the partition function of the replicated, constrained boson
\begin{equation}
\mathcal{Z}(n) = \mathcal{Z}^{-n} \int_{con} \mathcal{D}[\phi^a] e^{- \sum_{a=1}^n S[\phi^a] }
\end{equation}
for a compact boson of radius $R$ and action
\begin{equation}
S[\phi] = \frac{g}{4 \pi}\int d^2 x \, \, (\nabla \phi)^2 \, \, .
\end{equation}
We assume that the entanglement cut is a connected curve $\gamma$; the disconnected case (for a subsystem of two disks, for example, or an annulus) can be solved in a similar manner. In the connected case, the constraint takes the form $\phi^a(x) = \phi^b(x) + 2 \pi R \, w^{ab}$ along the cut, where $w^{ab} \in \mathbb{Z}$ are a set of relative winding numbers.  There is necessarily some ambiguity in the normalization of this constraint, leading to an extensive non-universal contribution $F(n) \sim \frac{L}{a}(n-1)$, which, for our purposes, can be ignored.

We assume an arbitrary mix of Dirichlet and Neumann external boundary conditions, suitably generalized to the compact case. The field configurations can be decomposed according to $\phi^a = \phi_{\alpha^a} + \tilde{\phi}^a$, where $\nabla^2 \phi_\alpha = 0$ encodes the vortices and zero modes  and $\tilde{\phi}$ is a fluctuation. Both obey the external boundary conditions. The fluctuations are \emph{not} compactified. Compactification affects only the topological sector, with $\alpha$ indexing whatever winding numbers or zero modes are required.
	
	The action $S[\phi] = S[\phi_\alpha] + S[\tilde{\phi}]$ decouples into the topological and fluctuating part, so the replicated partition function becomes
	
\begin{equation}
 \mathcal{Z}(n) = \mathcal{Z}^{-n} \sum_{\alpha^a: con} e^{-\sum S[\phi_{\alpha^a}]} \int_{{con}} \mathcal{D}[ \tilde{\phi}^a] e^{- \sum S[\tilde{\phi}^a] }
 \end{equation}
	The constraint imposes itself both in the topological and fluctuating sectors. For a fixed topological configuration $\alpha^a$, the fluctuations are constrained along the cut according to $\tilde{\phi}^a(x) - \tilde{\phi}^b(x)  = \phi_{\alpha^b}(x) - \phi_{\alpha^a}(x)  + 2 \pi R w^{ab}$. The relative winding numbers $w$ must also be summed over.
 
	 For fixed $\alpha^a$ and $w^{ab}$, we have a linear constraint on the fluctuations of the form $M \cdot \tilde{\phi}(x) = c(x)$ with less than full rank: the `center of mass' mode $\frac{1}{\sqrt{n}} \sum_a \tilde{\phi}^a$ is free. There are many possible choices of $M$ and $c$; for example, we can choose a cyclic form
\begin{equation}
M \cdot \tilde{\phi} = \left(\begin{array}{ccccc}-1 & 1 & 0 &  & 0 \\0 & -1 & 1 & \cdots & 0 \\0 & 0 & -1 &  & 0 \\ & \vdots &  & \ddots &   \\1 & 0 & 0 &  & -1\end{array}\right)\cdot \tilde{\phi} =  \left(\begin{array}{c}\phi_{\alpha^1}(x) - \phi_{\alpha^2}(x)  + 2 \pi R w^1 \\ \phi_{\alpha^2}(x) - \phi_{\alpha^3}(x)  + 2 \pi R w^2  \\ \vdots \\ \vdots \\ \phi_{\alpha^n}(x) - \phi_{\alpha^1}(x)  + 2 \pi R w^n \end{array}\right)
\end{equation}
with relative winding numbers $\vec{w} \in \mathbb{Z}^n : (1, 1, \cdots, 1) \cdot \vec{w} = 0 $. For any choice of $M$, $M$ has a right null vector $(1, 1, \cdots, 1)$ corresponding to the `center of mass mode', and a left null vector we call $d$. For the above choice of $M$, $d = (1, 1, \cdots, 1)$. Note that while a particular choice of $M$ may appear to break the permutation symmetry, the imposed constraint does not.

	As the fluctuations are not compactified, we can choose some $O(n)$ rotation $\tilde{\phi}' = T \tilde{\phi}$ such that $\tilde{\phi}^{'1}$ is the free `center of mass' mode. The measure and action $\mathcal{D}[ \tilde{\phi}^a] e^{- \sum S[\tilde{\phi}^{a}]} \to \mathcal{D}[ \tilde{\phi}^{'a}] e^{- \sum S[\tilde{\phi}^{'a}]}$ are left unchanged, but the constraint becomes
	 
\begin{equation}
M \cdot \tilde{\phi}(x) = c(x) \Rightarrow T M T^{-1} \cdot \tilde{\phi}^{'}(x) = T \cdot c(x)
\end{equation}
	 
	By our choice of $T$, the center of mass $\tilde{\phi}^{'1}$ does not appear in the constraint condition, so will contribute the partition function of unconstrained fluctuations which we denote by $\mathcal{Z}_f$. The remaining $(n-1)$ fields are constrained according to $\tilde{\phi}'(x) = T M^{-1} c(x) \equiv c'(x)$, as $M$ can be inverted in the subspace spanned by $c$. 
	
	 The precise form of $c'(x)$ is irrelevant, except that it is given by a fixed $\emph{linear}$ combination of the original $c(x)$. There is in addition a non-universal contribution $\sqrt{n}^{-L/a}$ to the partition function arising from careful consideration of the Jacobian of $T$ acting on the original constraint:

\begin{equation}
 \mathcal{Z}(n) = \frac{\mathcal{Z}_f }{\mathcal{Z}^n } \sqrt{n}^{-L/a}  \sum_{\alpha^a, w: con} e^{-\sum S[\phi_{\alpha^a}]} \int_{\tilde{\phi}'|_{\gamma} = c'(x)} \mathcal{D}[ \tilde{\phi}^{'a}] e^{- \sum_{a = 2}^n S[\tilde{\phi}^{'a}] }.
 \end{equation} 
	 
	 Again, we will ignore the non-universal extensive contribution. Finally, we use the shift invariance of the path integral to reduce the problem to Dirichlet constraints. Let ${C'}^a(x)$, defined on the entire region, satisfy the external boundary conditions as well as $C'(x)|_{\gamma} = c'(x)$ along the cut. Away from the cut we require $\nabla^2 C'(x) = 0$, which uniquely specifies the ${C'}^a$. We now shift $\tilde{\phi}' \to \tilde{\phi}' + C'$. The redefined fluctuations obey Dirichlet conditions along the cut so the action again decouples, 
\begin{equation}
S[\tilde{\phi}' + C'] = S[\tilde{\phi}'] + S[C'] 
\end{equation}	
The partition function becomes
\begin{equation}
\mathcal{Z}(n) = \frac{\mathcal{Z}_f }{\mathcal{Z}^n } \sum_{\alpha_a, w: con} e^{-\sum_{a = 1}^n S[\phi_{\alpha_a}] -  \sum_{a = 2}^n S[{C'}^a] } 
 \times \int_{\tilde{\phi}'|_{\gamma} = 0} \mathcal{D}[ \tilde{\phi}^a]_{} e^{- \sum_{a = 2}^{n}S[\tilde{\phi}^{'a}] } 
\end{equation}
The $(n-1)$ fluctuations decouple entirely from the topological sector, and are subject to Dirichlet constraints on the cut as well as the external boundary conditions. Let $\mathcal{Z}_D$ denote the partition function of the bosons obeying Dirichlet conditions on the cut, so that the fluctuations contribute $\mathcal{Z}_D^{n-1}$. Furthermore, the single replica partition function also decouples into a fluctuating and topological component, $\mathcal{Z} = \mathcal{Z}_f \cdot  \mathcal{Z}_{top}$. The full partition function is

\begin{equation}
\mathcal{Z}(n) = \left( \frac{\mathcal{Z}_D }{\mathcal{Z}_f} \right)^{n-1} \frac{1}{\mathcal{Z}_{top}^n}\sum_{\alpha_a: con} \sum_{ w } e^{- \sum_{a=1}^n S[\phi_\alpha] - \sum_{a=2}^n S[C'] } 
\end{equation}
The crucial observation is that $S[C']$ and $S[\phi_{\alpha}]$ are scale invariant and the sums require no UV regulation, so have exact conformal invariance. Any $L$ dependent terms of $\mathcal{Z}_{n}$ necessarily come from the fluctuations, $\left( \frac{\mathcal{Z}_D }{\mathcal{Z}_f} \right)^{n-1}$, which are identical to those of a non-compact boson. 	Thus, as far as the $\log(L)$ term is concerned, the rotation trick gives the correct result.

	To compute the $O(1)$ contributions, we must calculate $S[C']$. Define $C(x)|_{\gamma} = c(x)$ in the same manner as the $C'(x)$, which will be related according to $C'(x) = T M^{-1} C(x)$ (again, $M$ can be inverted in this subspace). All dependence on $T$ drops out, and we find

\begin{equation}
\sum_{a=2}^n S[C'] = \frac{g}{4 \pi} \int  \nabla C^T(x) (M M^T)^{-1} \nabla C(x)
\end{equation}
	
With knowledge of the classical solutions $C(x)$ for a given sector $\alpha, w$, the $O(1)$ term can then be computed.

\subsection{O(1) Terms for the Cylinder Geometry}
	We now reproduce the $O(1)$ contribution $\log(\sqrt{2 g} R) - \frac{1}{2}$ \cite{OshikawaS, HsuFradkin2S, StephanS} on an infinite cylinder. We consider a cylinder of circumference $L$ and length $2 \beta$, with the entanglement cut at $\tau = 0$. We assume Dirichlet conditions at $\tau = \pm \beta$, so that there are no topological indices $\alpha$ to sum over.
	
	A simple representation of the constraint is
	
\begin{equation}
M \cdot \tilde{\phi} = \left(\begin{array}{ccccc}0 & 0 & 0 &  & 0 \\-1 & 1 & 0 & \cdots & 0 \\-1 & 0 & 1 &  & 0 \\ & \vdots &  & \ddots &  0 \\-1 & 0 & 0 &  & 1\end{array}\right)\cdot \tilde{\phi} =  \left(\begin{array}{c}2 \pi R w^1 \\  2 \pi R w^2  \\ \vdots \\ \vdots \\ 2 \pi R w^n \end{array}\right)
\end{equation}

Obviously $w^1 = 0$, corresponding to the left null vector $d = (1, 0, \cdots, 0)$ of $M$. It is simple to find the desired classical solutions, by choosing $C(\tau, x) = c ( 1 - \frac{|\tau|}{\beta})$, with $c = 2 \pi R w$. We find

\begin{equation}
\sum_{a=2}^n S[C'] = \frac{g}{4 \pi} \frac{1}{\beta^2} (2 \pi R)^2 \int  w^T (M M^T)^{-1} w = \frac{g}{2 \pi} (2 \pi R)^2  \frac{L}{\beta} w^T (M M^T)^{-1} w
\end{equation}

Inserting into our general result, we have

\begin{align}
\mathcal{Z}(n) = \left( \frac{\mathcal{Z}_D }{\mathcal{Z}_f} \right)^{n-1} \sum_{ \vec{w} \in \mathbb{Z}^n : \vec{d}.\vec{w} = 0 } e^{ - \frac{g}{2 \pi} \frac{L}{\beta} (2 \pi R)^2 \vec{w}^T (M M^T)^{-1} \vec{w} } 
\end{align}

The sum is a Riemann Theta function, which we will evaluate in the large $\beta$ limit. The lattice sum goes over to an integral 

\begin{equation}
\sum_{ \vec{w} \in \mathbb{Z}^n : \vec{d}.\vec{w} = 0 } \to  \int dw^{n-1}
\end{equation}
where $dw^{n-1}$ is the Euclidean measure in the subspace $d.w = 0$. Standard gaussian integral identities give

\begin{equation}
\mathcal{Z}(n) = \left( \frac{\mathcal{Z}_D }{\mathcal{Z}_f} \right)^{n-1} \left(\frac{\beta}{L 2 g R^2}\right)^{(n-1)/2} \sqrt{|M M^T|'}
\end{equation}
where $|M M^T|'$ is the determinant of $M M^T$ with the zero mode omitted. We find $|M M^T|' = n$. So

\begin{equation}
\mathcal{Z}(n) = \left( \frac{\mathcal{Z}_D }{\mathcal{Z}_f} \right)^{n-1} \sqrt{n} \, \left(\frac{\beta}{L}\right)^{(n-1)/2} (\sqrt{2 g} R)^{-(n-1)} 
\end{equation}

The asymptotics of $\mathcal{Z}_{D/f}$ (which are given by $\eta$ functions) show that the fluctuating part does not contribute at $O(1)$. Furthermore, dependence on the modular parameter $\frac{\beta}{L}$ contributes an overall divergent constant, so can be ignored. We extract the universal part

\begin{equation}
\mathcal{Z}(n) =  (\sqrt{2 g} R)^{-(n-1)} \sqrt{n}
\end{equation}
with resultant entanglement entropy $S =\log( \sqrt{2 g} R ) - \frac{1}{2}$ as desired.

\end{document}